\documentclass[10pt,twocolumn,showpacs,titlepage,aps]{revtex4}%
\usepackage{amsfonts}
\usepackage{amsmath}
\usepackage{amssymb}
\usepackage{graphicx}%
\providecommand{\U}[1]{\protect \rule{.1in}{.1in}}
\begin{document}
\title{Magnetic rogue wave in a perpendicular anisotropic ferromagnetic nanowire with
spin-transfer torque }
\author{Fei Zhao$^{1}$, Zai-Dong Li$^{1}$, }
\thanks{Corresponding author, E-mail: zdli2003@yahoo.com}
\author{Qiu-Yan Li$^{1}$, Lin Wen$^{2}$, Guangsheng Fu$^{1}$, W. M. Liu$^{2}$}
\affiliation{$^{1}$Department of Applied Physics and School of Information Engineering,
Hebei University of Technology, Tianjin 300401, China}
\affiliation{$^{2}$Beijing National Laboratory for Condensed Matter Physics, Institute of
Physics, Chinese Academy of Sciences, Beijing 100080, China}

\begin{abstract}
We present the current controlled motion of dynamic soliton embedded in spin
wave background in ferromagnetic nanowire. With the stronger breather
character we get the novel magnetic rogue wave and clarify its formation
mechanism. The generation of magnetic rogue wave is mainly arose from the
accumulation of energy and magnons toward to its central part. We also observe
that the spin-polarized current can control the exchange rate of magnons
between envelope soliton and background, and the critical current condition is
obtained analytically. Even more interesting is that the spin-transfer torque
plays the completely opposite role for the cases of below and above the
critical value.

\end{abstract}

\maketitle

\section{Introduction}

In ferromagnet nanowires, the deviation of magnetization from the ground state
results in the excitation of spin waves. Their attractive interaction and
instabilities contribute to the existence of topological and dynamic solitons.
The dynamics of magnetization in a soliton can be well described by the famous
Landau-Lifshitz-Gilbert equation \cite{Gilbert}.

The topological soliton, now namely domain wall (DW), connecting two vacua
describes an inhomogeneous state of magnetization and it cannot be reduced to
the ground state by any finite deformation \cite{Kose}. So it has
technological application in race track memory \cite{Parkin}. The field-driven
DW motion has been studied extensively in thin films \cite{Hub} and recently
in magnetic nanowires \cite{Ono,Beach,Hay,Thiaville,Grollier,Mar}. These
studies show that the DW travelling speed in magnetic materials conventionally
depends on the strength of a constant magnetic field as the external filed is
below the Walker-breakdown value \cite{Schryer,Malo}. Above this critical
value, the DW develops a variational complex internal structure which results
in temporal oscillation of DW velocity \cite{Ono,Beach,Beach1,Wang}. Recently,
many studies demonstrate that the spin-polarized current can cause many unique
phenomena for magnetization motion, which attributes to spin-transfer effect
\cite{Slonczewski}. With this consideration the modified
Landau-Lifshitz-Gilbert equation \cite{Bazaliy,Tatara,SZhang} including
spin-transfer torque has been derived to describe such current-induced
magnetization dynamics. With the remarkable experimental successes measuring
the motion of a DW, considerable progresses have been made to understand the
current-induced DW motion in magnetic nanowires
\cite{Tatara,SZhang,Yama,Sai,Lim,Ohe,li2010}. This current induced
magnetization motion possesses of different character from that driven by an
magnetic field, and the spin-transfer torque can induce not only precession
but also damping role for the motion of magnetization.

On the other hand, the dynamic soliton describes the localized states of
magnetization which can be reduced to a uniform magnetization by continuous
deformation, so that the excited ferromagnet makes a transition to the ground
state. Therefore, a dynamic soliton is sometimes said to be topologically
equivalent to the ground state. Also, the motion of dynamic soliton is of
topic research in confined ferromagnetic materials
\cite{Belliard,Yamada,Tsoi1,Linear}, especially with the generation and
detection of magnons excitation \cite{Tsoi} in a magnetic multilayer. Driven
by the adiabatic spin-transfer torque, the dynamic soliton solutions for
isotropic case \cite{Lizd} and uniaxial anisotropic case \cite{lizd2007} are
investigated carefully, where the corresponding solutions show the
characteristic breather behavior for magnetization motion in ferromagnetic
nanowire. However, the current driven motion of the dynamic soliton are not
well explored carefully.

In this paper, we present the generation of a novel \textit{magnetic rouge
wave} solution under stronger breather characters in a perpendicular
anisotropic ferromagnetic nanowire driven by spin-transfer torque, which is
similar to optical rouge wave in fiber \cite{Shu, Kib}. The formation
mechanism of magnetic rouge wave is clarified carefully. In this process, we
observe that the spin-polarized current can control the magnons exchange rate
between the envelope soliton and background, and the critical current
condition is obtained analytically. Moreover, the spin-transfer torque play
the completely opposite role for the cases of below and above the critical value.

\section{Current driven interaction of spin wave and a dynamic soliton in
ferromagnetic nanowire}

When a spin-polarized electric current flows the ferromagnetic nanowire, the
localized magnetization dynamics can be described by the modified
Landau-Lifshitz equation \cite{Tatara,SZhang} with the spin-transfer torque,%
\begin{equation}
\frac{\partial \mathbf{M}}{\partial t}=-\gamma \mathbf{M}\times \mathbf{H}%
_{\text{eff}}+\frac{\alpha}{M_{s}}\mathbf{M\times}\frac{\partial \mathbf{M}%
}{\partial t}+\mathbf{\tau}_{s}, \label{LLG}%
\end{equation}
where $\gamma$ is the gyromagnetic ratio, $\alpha$ is the Gilbert damping
parameter, and $\mathbf{H}_{\text{eff}}$ represents the effective magnetic
field including the exchange field, the anisotropy field and the external
field. For a perpendicular anisotropic ferromagnetic nanowire, the effective
magnetic field takes the form $\mathbf{H}_{\text{eff}}=\left(  2F/M_{s}%
^{2}\right)  \partial^{2}\mathbf{M/}\partial x^{2}+[\left(  H_{K}/M_{s}%
-4\pi \right)  M_{z}+H_{\text{ext}}]\mathbf{e}_{z}$, where is the exchange
constant, $H_{K}$ denotes the energetic anisotropy coefficient, $H_{\text{ext}%
}$ is the applied external magnetic field, and $\mathbf{e}_{z}$ is the unit
vector along the $z$-axis. The last term $\mathbf{\tau}_{s}$ in Eq.
(\ref{LLG}) denotes the spin-transfer torque, which describes the injected
current can be polarized and produces a torque acting on the local
magnetization. This adiabatic spin-transfer torque is of the form
$\mathbf{\tau}_{s}=b_{J}(\partial M/\partial x)$ \cite{SZhang}, where
$b_{J}=Pj_{e}\mu_{B}/\left(  eM_{s}\right)  $, $P$ is the spin polarization of
the current, $j_{e}$ is the electric current density and flows along the $x$
direction, $\mu_{B}$ is the Bohr magneton, $e$ is the magnitude of electron charge.

Introducing the normalized magnetization, i.e., $\mathbf{m}=\mathbf{M}/M_{s}$,
we can simplify Eq. (\ref{LLG}) as the dimensionless form%
\begin{equation}
\frac{\partial \mathbf{m}}{\partial t}=-\mathbf{m}\times \mathbf{h}_{\text{eff}%
}+\alpha \mathbf{m\times}\frac{\partial \mathbf{m}}{\partial t}+A_{J}%
\frac{\partial \mathbf{m}}{\partial x}, \label{LLG1}%
\end{equation}
where $\mathbf{h}_{\text{eff}}=\mathbf{H}_{\text{eff}}/M_{s}$ and $A_{J}%
=b_{J}t_{0}/l_{0}$. The time $t$ and space coordinate $x$ have been rescaled
by the characteristic time $t_{0}=1/\left[  \gamma \left(  H_{K}-4\pi
M_{s}\right)  \right]  $ and length $l_{0}=\sqrt{2F/\left[  \left(  H_{K}-4\pi
M_{s}\right)  M_{s}\right]  }$, respectively. As discussed in our pervious
work \cite{lizd2007}, Eq. (\ref{LLG1}) admits the excited states, i.e., spin
wave and a dynamic soliton. For the perpendicular anisotropic ferromagnetic
nanowire, these two types of excited state denote the small deviation of
magnetization from the ground state. Therefore, a reasonable complex function
$q$ can be introduced to replace the components of normalized magnetization
\cite{Kose}, i.e., $q\equiv m_{x}+im_{y}$ and $m_{z}^{2}=1-\left \vert
q\right \vert ^{2}$. Under the long-wavelength approximation \cite{Kose} and
without damping, Eq. (\ref{LLG1}) becomes the integrable nonlinear
Schr\"{o}dinger equation%
\begin{equation}
i\frac{\partial q}{\partial t}=\frac{\partial^{2}q}{\partial x^{2}}+\frac
{1}{2}q\left \vert q\right \vert ^{2}+iA_{J}\frac{\partial q}{\partial x}%
-\omega_{0}q, \label{nls}%
\end{equation}
where $\omega_{0}=1+H_{\text{ext}}/\left(  H_{k}-4\pi M_{s}\right)  $.

\begin{figure}[ptb]
\includegraphics[width=8cm]{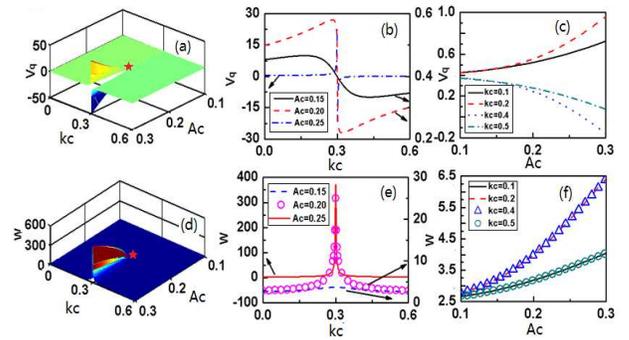}\caption{(Color online) Breather velocity
$V_{q}$ and width $W$ vs the amplitude $A_{c}$ and wave number $k_{c}$ of spin
wave. The other parameters are $u_{1}=0.2$, $\nu_{1}=-0.3$ and $A_{J}=0.2$.
(a) and (d): The soliton velocity $V_{q}$ and width $W$ vs with $A_{c}$ and
$k_{c}$. The red star signs the location of the rogue wave. (b) and (c) The
dependence of velocity on $A_{c}$ and $k_{c}$. (e) and (f) The dependence of
velocity on $A_{c}$ and $k_{c}$.}%
\end{figure}

It is easy to find two basic solutions of Eq. (\ref{nls}). One is $q=0$, which
corresponds to the ground state $\mathbf{m}=\left(  0,0,1\right)  $. The other
solution is spin wave, i.e., $q=A_{c}e^{-i\left(  k_{c}x-\omega_{c}t\right)
}$ with $\omega_{c}$ and $k_{c}$ being the dimensionless frequency and wave
number, respectively. In perpendicular anisotropic ferromagnetic nanowire, the
interaction of spin waves is attractive and their instabilities lead to
macroscopic phenomena, i.e., the appearance of a spatially localized magnetic
excited state (topological or dynamic soliton). As shown later, the
interaction of spin wave and a dynamic soliton can result in the \textit{novel
magnetic rogue wave} which can be realized by adjusting the relation of wave
number and amplitude for spin wave and a dynamic soliton. To this purpose, one
should get the solution on the spin wave background.

By employing Darboux transformation \cite{lizd2007,Matveev} and performing a
tedious calculation, we obtain such solution
\begin{equation}
q=e^{i\varphi}[A_{c}+2u_{1}\left(  \Delta_{1}+i\Delta_{2}\right)  /\Delta],
\label{sol1}%
\end{equation}
where $\varphi=\omega_{c}t-k_{c}x$, $\Delta=\cosh \theta+a\cos \beta$,
$\Delta_{1}=a\cosh \theta+\cos \beta$, $\Delta_{2}=b\sinh \theta+c\sin \beta$,
here $\theta=D_{R}x+(D\delta)_{R}t+x_{0}$, $\beta=D_{I}x+\left(
D\delta \right)  _{I}t-t_{0}$, $a=2A_{c}L_{R}/(\left \vert L\right \vert
^{2}+A_{c}^{2})$, $b=2A_{c}L_{I}/(\left \vert L\right \vert ^{2}+A_{c}^{2})$,
$c=(\left \vert L\right \vert ^{2}-A_{c}^{2})/(\left \vert L\right \vert
^{2}+A_{c}^{2})$, and the subscript $\allowbreak R$ and $\allowbreak I$
represent the real and imaginary part, respectively. The other parameters are
$L=-ik_{c}-D-\lambda$, $D=[\left(  ik_{c}+\lambda \right)  ^{2}-A_{c}%
^{2}]^{1/2}$, and $\delta=-i\lambda-k_{c}+A_{J}$, where $\lambda=u_{1}%
+i\nu_{1}$. Here $x_{0}$, $t_{0}$, $A_{c}$, $k_{c}$, $u_{1}$, and $\nu_{1}$
are the real constants, and without loss of generality we assume that $A_{c}$
and $u_{1}$ are non-negative real constants.\emph{ }

The solution in Eq. (\ref{sol1}) exhibits a dynamic soliton solution embedded
in spin wave background. Two types of excited states (spin wave and a dynamic
soliton) can be recovered from Eq. (\ref{sol1}): a) As the spin wave amplitude
and wave number vanishes, namely $A_{c}=k_{c}=0$, we can get a dynamic
soliton, $q_{1}=2u_{1}e^{i\left(  \beta_{1}+\omega_{c}t\right)  }/\cosh
\theta_{1}$, with $\theta_{1}=u_{1}x+u_{1}\left(  2\nu_{1}+A_{J}\right)
t+x_{0}$ and $\beta_{1}=\nu_{1}x-\left(  u_{1}^{2}-\nu_{1}^{2}-A_{J}\nu
_{1}\right)  t-t_{0}$. This solution corresponds to the dynamic precession of
magnetic soliton on the ground state background, where the soliton moves with
the velocity $v=-2\nu_{1}-A_{J}$, and the components $m_{x}$ and $m_{y}$
precess around $m_{z}$ with the amplitude $A_{q}=2u_{1}/\cosh \theta_{1}$ and
the frequency $\omega_{q}=u_{1}^{2}-\nu_{1}^{2}-A_{J}\nu_{1}+\omega_{c}$. It
clearly demonstrates that the spin current term $A_{J}$ can change the dynamic
soliton velocity and the precessional frequency. In addition, the solution
$q_{1}$\ represents in fact the static magnetic soliton with three integrals
of the motion and the uniform\ magnon density $\left \vert q_{1}\right \vert
^{2}=$ $2u_{1}$ along the direction of soliton propagation. b) When the
amplitude of dynamic soliton vanishes, namely $u_{1}=0$, Eq. (\ref{sol1})
reduces to spin wave solution $q=A_{c}e^{-i\left(  k_{c}x-\omega_{c}t\right)
}$, in which the magnon density is also constant.

\begin{figure}[ptb]
\includegraphics[width=8.5cm]{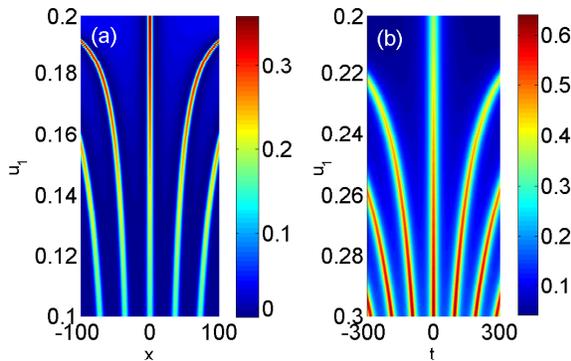}\caption{(Color online) The evolution
of magnetic soliton in the limit processes $u_{1}\rightarrow A_{c}^{-}$ (a)
and $u_{1}\rightarrow A_{c}^{+}$ (b). Other parameters are $A_{c}=0.2$,
$A_{J}=0.2$, $k_{c}=-\nu_{1}=0.003$, and $x_{0}=t_{0}=0$.}%
\end{figure}

From Eq. (\ref{sol1}), we observe that the solution commonly exhibits a
breather character and a time periodic modulation for soliton amplitude, which
in fact denotes the interaction between the localized process of spin wave
background and the periodization process of magnetic soliton. The properties
of the soliton solution are characterized by the slope direction $V_{\theta
}\equiv D_{R}x+(D\delta)_{R}t$, the breather propagation velocity $v_{q}%
\equiv-(D\delta)_{R}/D_{R}$ and the soliton width $1/D_{R}$. With the
expressions of $D$ and $\delta$, we find that the soliton velocity and width
are modulated by the amplitude $A_{c}$ and wave number $k_{c}$ of spin wave as
shown in Fig. 1(a) and 1(d). The absolute value of soliton velocity and
soliton width increases with increasing $A_{c}$ as shown in Fig. 1(c) and
1(f). Moreover, as the modulation parameter $A_{c}$ increases continuously,
the soliton velocity will be affected mainly by the value of $k_{c}$ near
$-\nu_{1}$. When $k_{c}=-\nu_{1}$, the soliton velocity and width is maximal,
respectively, as shown in Fig. 1(b) and 1(e). This phenomena in fact ascribes
to magnonic spin-transfer torque \cite{Yanp} which denotes the transfer of
spin angular momentum from spin wave background to a dynamic soliton, and it
will be discussed elsewhere in detail. On the basis of Eq. (\ref{sol1}), as
the modulation parameter $u_{1}$ approaches $A_{c}$, some novel properties
will be presented. Magnetic rouge wave solution will be excited in
ferromagnetic nanowire which leads to some fantastic phenomena.

\section{Novel magnetic rogue wave}

In terms of our previous discussion \cite{lizd2007} for the solution in Eq.
(\ref{sol1}), we have known\ that the critical point $u_{1}=A_{c}$ forms a
dividing line between the modulation instability process $\left(  u_{1}%
>A_{c}\right)  $ and the periodization process $\left(  u_{1}<A_{c}\right)  $
under the condition $\nu_{1}=-k_{c}$. It leads to the different physical
behavior how the breather character depends strongly on the modulation
parameter $u_{1}$ as shown in Fig. 2. Fig. 2 shows two different asymptotic
behavior in the limit processes $u_{1}\rightarrow A_{c}^{-}$ (and $A_{c}^{+}$)
under the condition $\nu_{1}=-k_{c}$, respectively. The former demonstrates a
spatial periodic process of a soliton and we observe that the spatial
separation of adjacent peak and each peak value increases rapidly as the
modulation parameter $u_{1}$ approaching $A_{c}$, respectively. The latter
shows a localized process of the spin-wave background along the slope
direction $K_{\beta}=-(D\delta)_{I}/D_{I}$ for $t_{0}=0$. In this case, the
temporal separation of adjacent peak also increases rapidly as the modulation
parameter $u_{1}$ approaching $A_{c}$, while each peak value takes the rapid
decrease. Even more interesting, in the limit case of $u_{1}\rightarrow A_{c}%
$, i.e., the pentagram sign indicated in Fig 1(a) and (d), we get the
\textit{novel magnetic rogue wave}
\begin{equation}
Q_{1}=A_{c}e^{i\varphi}\left[  \frac{4\left(  1-itA_{c}^{2}\right)  }%
{t^{2}A_{c}^{2}\eta+2txA_{c}^{2}\zeta+\varepsilon}-1\right]  , \label{sol2}%
\end{equation}
where $\eta=$ $A_{J}^{2}+A_{c}^{2}+4k_{c}^{2}-4A_{J}k_{c}$, $\zeta
=A_{J}-2k_{c}$, and $\varepsilon=1+x^{2}A_{c}^{2}\allowbreak$. Eq.
(\ref{sol2}) shows the typical rogue wave feature that the magnons accumulated
from spin wave background converge a single hump with the critical amplitude
$A_{Q}=3A_{c}$. It implies that the localization wave is captured completely
at $x=0$ and $t=0$ by spin wave background. As shown in Fig. 3, the
realization of magnetic rogue wave in Eq. (\ref{sol2}) attributes to that the
aggregation of magnons gradually increases as $u_{1}$ approaches $A_{c}^{-}$,
while decreases gradually with $u_{1}$ approaches $A_{c}^{+}$. The temporal
localization magnetic rogue wave is excited as $u_{1}\rightarrow A_{c}$\ with
the magnon density peak $\left \vert q\right \vert ^{2}=9A_{c}^{2}$.

\begin{figure}[ptb]
\begin{center}
\includegraphics[width=7.5cm]{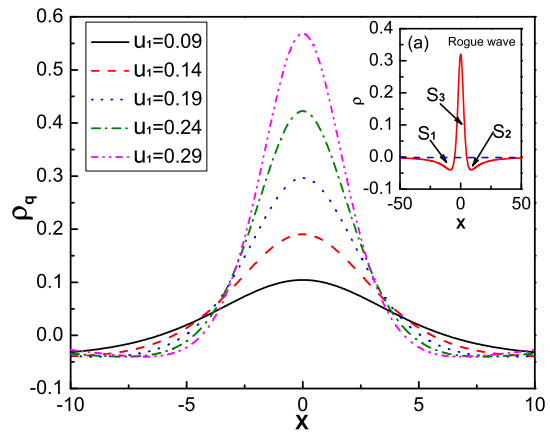}
\end{center}
\caption{(Color online) The magnon density distribution against the background
for the different parameter $u_{1}$, which ranges from $0.09$ to $0.29$ in
$0.05$ steps. (a) The magnon density distribution against the background for
the excited formation of magnetic rogue wave. Other parameters are $A_{c}%
=0.2$, $A_{J}=k_{c}=0.1$ and $x_{0}=t_{0}=0.$}%
\end{figure}

In order to investigate deeply the properties of magnetic rogue wave in
ferromagnetic nanowire, the analysis for the magnon density distribution
against the background plays a major role, which is defined by the quantity
$\rho_{q}\left(  x,t\right)  =\left \vert q_{1}\left(  x,t\right)  \right \vert
^{2}-\left \vert q_{1}\left(  x=\pm \infty,t\right)  \right \vert ^{2}$. In Fig.
3 we plot the evolution of magnon density distribution for the breather
solution in Eq. (\ref{sol1}), which ultimately demonstrates the formative
mechanism of magnetic rogue wave solution in ferromagnetic nanowire. With the
modulation parameter $u_{1}$ approaches $A_{c}$, the magnons in the background
gradually gather toward to each individual central part and the envelope
becomes sharper. Specially, as shown the inset (a), the critical peak will
appear under the condition $u_{1}\rightarrow A_{c}$, i.e., the magnetic rogue
wave is excited.

With Eq. (\ref{sol2}) and the quantity $\rho=\left \vert Q_{1}\left(
x,t\right)  \right \vert ^{2}-\left \vert Q_{1}\left(  x=\pm \infty,t\right)
\right \vert ^{2}$ we obtain the magnon density distribution in a magnetic
rogue wave
\begin{equation}
\rho=8A_{c}^{2}\frac{\Gamma_{1}-\Gamma_{2}}{\left(  \Gamma_{1}+\Gamma
_{2}\right)  ^{2}}, \label{sol3}%
\end{equation}
where $\Gamma_{1}=1+t^{2}A_{c}^{4}$, $\Gamma_{2}=A_{c}^{2}\left(  x+t\left(
A_{J}-2k_{c}\right)  \right)  ^{2}$, and Eq. (\ref{sol3}) implies the integral
$\int_{-\infty}^{+\infty}\rho(x,t)dx=0$ for arbitrary time. From the condition
$\rho_{Q}\left(  \pm1/A_{c},0\right)  =0$, we can define the spatial width of
the hump part in rogue wave as $2/A_{c}$. A detail calculation shows that at a
fixed time the loss of magnons in background completely transfer to hump,
i.e., the area relation $\mathbf{S}_{1}+\mathbf{S}_{2}=\mathbf{S}_{3}$. These
results clearly illustrate that the generation of magnetic rogue wave with
stronger breather character is mainly arose from the gathering energy and
magnons from the background toward to its central part, and the loss of
magnons in background completely transfer to the hump part in magnetic rogue
wave. \begin{figure}[ptb]
\includegraphics[width=7.5cm]{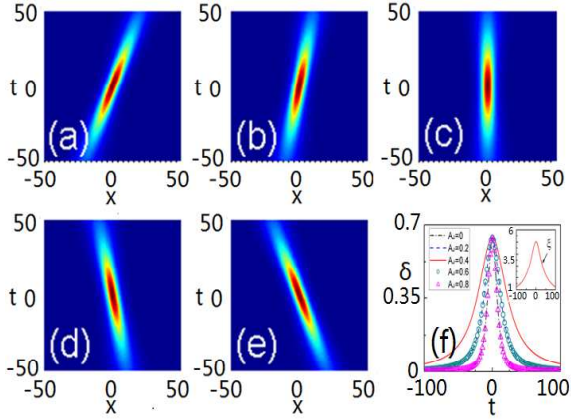}\caption{(Color online) (a)-(e) The
formation region in space ($x,t$) for magnetic rogue wave with different
current. The parameter $A_{J}$\ ranges form $0$ to $0.8$ in $0.2$ steps. (f)
The nonuniform exchange of magnons between rogue wave and background for the
different spin current. The inset figure in (f) denotes the maximal
accumulation (or dissipation) process for the critical current value
$A_{J}=2k_{c}$. Other parameters are $A_{c}=0.2$ and $k_{c}=0.2$.}%
\end{figure}

The other interesting fundamental problem is that how rogue wave gather
magnons and energy toward to its central part from the background. This can be
explained by the quantity $\delta \left(  x,t\right)  \equiv \lim_{l_{Q}%
\rightarrow \pm \infty}\left \vert Q_{1}\left(  x,t\right)  \right.  -\left.
Q_{1}\left(  x=l_{Q},t\right)  \right \vert ^{2}$. With Eq. (\ref{sol2}) we
obtain%
\begin{equation}
\delta \left(  x,t\right)  =16A_{c}^{2}\frac{\Gamma_{1}}{\left(  \Gamma
_{1}+\Gamma_{2}\right)  ^{2}}, \label{sol4}%
\end{equation}
which denotes the nonuniform exchange of magnons between rogue wave and
background for the different spin current as shown in Fig. 4. From Eq.
(\ref{sol4}) we find the spin current can control the accumulation and
dissipation rate of magnons, and there is a critical current condition, i.e.,
$A_{Jc}=2k_{c}$. Below the critical current, the magnons exchange decreases
with the increasing current term $A_{J}$. However, the magnons exchange is
accelerated with the increasing current above the critical value. The roles of
spin-transfer torque are completely opposite for the cases below and above the
critical current which is shown in Fig. 4 (f). From Fig. 4(a) to 4(e) we see
that the magnetic rogue wave can be created in the different direction for
($x,t$) space, which ascribes to the nonuniform exchange of magnons between
rogue wave and background tuning by the spin-polarized current. When
$A_{J}=2k_{c}$, the time of magnons accumulation (or dissipation) attains its
maximum. As shown the inset figure of Fig. 4 (f), i.e., the integral
$\xi \left(  x,t\right)  =\int_{-\infty}^{+\infty}\delta \left(  x,t\right)
dx=8\pi A_{c}/(1+t^{2}A_{c}^{4})^{1/2}$, the magnons in background accumulate
to the central part when $t<0$. It leads to the generation of a hump with two
grooves on the background along the space direction and the critical peak of
the hump can occur at $t=0$. In contrast, when $t>0$, the magnons in the hump
start to dissipate into the background so that the hump gradually decay. The
magnetic rogue wave disappears ultimately, and it verifies the rogue wave is
only one oscillation in temporal localization and displays a unstable dynamic
behavior.\begin{figure}[ptb]
\includegraphics[width=8cm]{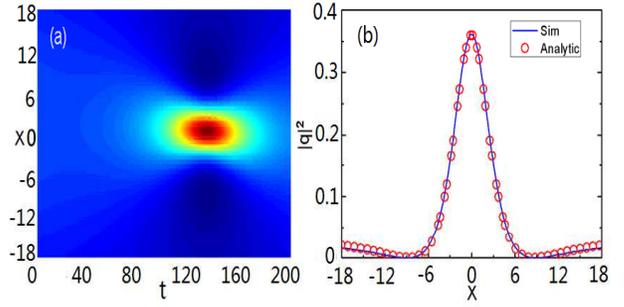}\caption{(Color online) (a) Evolution of
the numerical magnetic rogue wave. (b) Comparison of magnon density between
the numerical solution (in solid curve) and ideal solution (in '$\circ$').
Other parameter are $A_{c}=0.2$, $u_{1}=0.19$, $x_{0}=0.72,$ $A_{J}=0.01$,
$k_{c}=0.2$ and $\omega=0.2$.}%
\end{figure}

With some specific initial conditions, the excitation of rogue wave can be
recovered by means of the numerical simulation. In order to understand such
process of magnetic rogue wave, we choose the initial value of solution, which
can be approximated by%
\begin{equation}
q\left(  x,0\right)  =\left(  \rho+\epsilon \chi \cos \varphi_{1}\right)
\exp \left(  i\varphi \right)  , \label{init1}%
\end{equation}
where $\rho=\left[  2\kappa_{1}\left(  \kappa_{1}-iu_{1}\right)  -A_{c}%
^{2}\right]  /A_{c}$, $\epsilon=\exp \left(  -x_{0}\right)  $, $\chi
=4\kappa_{1}u_{1}\left(  \kappa_{1}-iu_{1}\right)  /A_{c}^{2}$ and
$\varphi_{1}=\kappa_{1}x$ with $\kappa_{1}=\sqrt{A_{c}^{2}-u_{1}^{2}}$. By
direct numerical simulations, we find that the solution of the initial value
problem in Eq. (\ref{nls}) with initial condition Eq. (\ref{init1}) can be
well described by the solution in Eq. (\ref{sol2}). Fig. 5(a) is a result of
direct numerical simulation, which shows that a small periodic perturbation
with a very small modulation parameter can induce a near-ideal rogue wave
localization, whose profile is basically consistent with the ideal theoretical
limit solution of Eq. (\ref{sol2}) as shown in Fig. 5(b). As a result, a small
initial perturbation with a small modulation can induce the generation and
breakup of a near-ideal rogue wave.

\section{Conclusions}

In summary, we have investigated the formative mechanism of magnetic rogue
wave and the properties of magnon density in uniaxial anisotropic
ferromagnetic nanowire driven by spin-transfer torque. Our results show that
the accumulation of energy and magnons toward to its central part plays the
main role for the generation of rogue wave with stronger breathing character
and rogue wave only appears at the spatial-temporal localization. We also
display the rouge wave is unsuitably and gradually decay because the
nonuniform exchange rate of energy and magnons which can be controlled by the
spin-polarized current. A novel critical current condition is obtained
analytically, and the spin-transfer torque plays the completely opposite roles
for the case of below and above the critical value.

\section{Acknowledgement}

Zai-Dong Li was supported by NSF of China under Grant No. 10874038, the
Hundred Innovation Talents Supporting Project of Hebei Province of China under
Grant No. CPRC014, Tianjin Municipal Natural Science Foundation of China (No.
11JCYBJC01600), and China postdoctoral science foundation under Grant No.
20100470987. This work was also supported by NSFC under grants Nos. 10874235,
10934010, 60978019, the NKBRSFC under grants Nos. 2009CB930701, 2010CB922904,
2011CB921502, 2012CB821300, and NSFC-RGC under grants Nos. 11061160490 and 1386-N-HKU748/10.

\end{document}